\documentclass[10pt,aps,prc,twocolumn,superscriptaddress,floatfix,nofootinbib]{revtex4-1}
\usepackage{amsmath,amssymb,mathrsfs}
\usepackage{bm}
\usepackage{ascmac}
\usepackage{geometry}
\usepackage{graphicx}

\def\0\\{\nonumber\\}
\def\bs#1{\boldsymbol{#1}}

\def\zI{\mathrm{i}\hspace{0.2mm}}
\def\fs#1{{\footnotesize #1}}

\makeatletter
\newcommand\footnoteref[1]{\protected@xdef\@thefnmark{\ref{#1}}\@footnotemark}
\makeatother

\begin{document}

\title{
Microscopic description of production cross sections including deexcitation effects
}

\author{Kazuyuki Sekizawa}
\email[]{sekizawa@if.pw.edu.pl}
\affiliation{Faculty of Physics, Warsaw University of Technology, ulica Koszykowa 75, 00-662 Warsaw, Poland}

\date{July 7, 2017}

\begin{abstract}
\begin{description}
\item[Background]
At the forefront of the nuclear science, production of new neutron-rich isotopes
is continuously pursued at accelerator laboratories all over the world. To explore
the currently-unknown territories in the nuclear chart far away from the stability,
reliable theoretical predictions are inevitable.

\item[Purpose]
To provide a reliable prediction of production cross sections taking into account secondary
deexcitation processes, both particle evaporation and fission, a new method called TDHF+GEMINI
is proposed, which combines the microscopic time-dependent Hartree-Fock (TDHF) theory with
a sophisticated statistical compound-nucleus deexcitation model, {\scriptsize GEMINI}++.

\item[Methods]
Low-energy heavy ion reactions are described based on three-dimensional Skyrme-TDHF
calculations. Using particle-number projection method, production probabilities, total angular
momenta, and excitation energies of primary reaction products are extracted from the TDHF
wavefunction after collision. Production cross sections for secondary reaction products are
evaluated employing {\scriptsize GEMINI}++. Results are compared with available experimental
data and widely-used \fs{GRAZING} calculations.

\item[Results]
The method is applied to describe cross sections for multinucleon transfer processes in
$^{40}$Ca+$^{124}$Sn ($E_{\rm c.m.}$\,$\simeq$\,128.54\,MeV),
$^{48}$Ca+$^{124}$Sn ($E_{\rm c.m.}$\,$\simeq$\,125.44\,MeV),
$^{40}$Ca+$^{208}$Pb ($E_{\rm c.m.}$\,$\simeq$\,208.84\,MeV),
$^{58}$Ni+$^{208}$Pb ($E_{\rm c.m.}$\,$\simeq$\,256.79\,MeV),
$^{64}$Ni+$^{238}$U ($E_{\rm c.m.}$\,$\simeq$\,307.35\,MeV), and
$^{136}$Xe+$^{198}$Pt ($E_{\rm c.m.}$\,$\simeq$\,644.98\,MeV) reactions at energies
close to the Coulomb barrier. It is shown that the inclusion of secondary deexcitation processes,
which are dominated by neutron evaporation in the present systems, substantially improves agreement
with the experimental data. The magnitude of the evaporation effects is very similar to the one observed
in \fs{GRAZING} calculations. TDHF+GEMINI provides better description of the absolute value of
the cross sections for channels involving transfer of more than one protons, compared to the
\fs{GRAZING} results. However, there remain discrepancies between the measurements and
the calculated cross sections, indicating a limit of the theoretical framework that works with
a single mean-field potential. Possible causes of the discrepancies are discussed.

\item[Conclusions]
In order to perfectly reproduce experimental cross sections for multinucleon transfer processes,
one should go beyond the standard self-consistent mean-field description. Nevertheless, the
proposed method will provide valuable information to optimize production mechanisms of new
neutron-rich nuclei through its microscopic, non-empirical predictions.

\end{description}
\end{abstract}


\maketitle

\section{INTRODUCTION}

To expand our knowledge of the nature of the atomic nuclei, it is obviously
important to produce new neutron-rich unstable isotopes that have not yet been
produced to date, and study their properties both experimentally and theoretically.
However, the optimal reaction condition, such as projectile-target combinations
and incident energies, to produce such extremely unstable nuclei is not obvious,
and reliable theoretical predictions are mandatory to guide experiments at current
and future radioactive-ion beam facilities. This paper aims to provide a predictive
model of production cross sections in low-energy heavy ion reactions.

To describe low-energy heavy ion reactions, various models have been developed.
Semi-classical models, called \fs{GRAZING} \cite{GRAZING} and complex
Wentzel-Kramers-Brillouin (CWKB) \cite{CWKB}, have shown remarkable
successes in describing multinucleon transfer (MNT) processes in peripheral
collisions \cite{Corradi(review)}. The \fs{GRAZING} code was extended to
include effects of transfer-induced fission in competition with particle evaporation
\cite{GRAZING-F}. A possible drawback of those models lies in insufficient
description of deep-inelastic processes at small impact parameters. On the other hand,
different theoretical approaches have also been developed: \textit{e.g.} a dynamical
model based on Langevin-type equations of motion \cite{Zagrebaev(2005),Zagrebaev(2007)1,
Zagrebaev(2008)1,Zagrebaev(2013)IQF,Zagrebaev(2014)light_N-rich,Zagrebaev(2008)2,
Zagrebaev(2007)2,EXP2015(136Xe+208Pb)}, dinuclear system model (DNS) \cite{Antonenko(1995),
Adamian(1997)1,Adamian(1997)2,Adamian(1998),Adamian(2003),Feng(2009),Adamian(2010)1,
Adamian(2010)2,Adamian(2010)3,Mun(2014),Mun(2015),DNS(2015),DNS(2017),DNS(2017)2,
DNS(2017)3,DNS(2017)4}, and improved quantum molecular dynamics model (ImQMD)
\cite{ImQMD(2002),ImQMD(2004),ImQMD(2008),ImQMD(2013),ImQMD(2015)1,ImQMD(2015)2,
ImQMD(2016)1,ImQMD(2016)2,ImQMD(2016)3,ImQMD(2017)}. Although those models
can describe both peripheral and damped collisions, including fusion and quasifission (QF)
processes, they are to some extent empirical containing model parameters. In order to
provide a reliable prediction to produce new neutron-rich isotopes, it is desirable to have
less adjustable parameters as possible. In the present paper, a method is developed to
predict production cross sections based on a microscopic framework of the time-dependent
Hartree-Fock (TDHF) theory.

The TDHF theory allows to describe nuclear dynamics microscopically from nucleonic
degrees of freedom. The theory itself was proposed in 1930 \cite{Dirac(TDHF)}, and
its application to nuclear systems already started about forty years ago \cite{BKN(1976),
Negele(review)}. Since then, it has been developed as an omnipotent tool, rooted with the
time-dependent density functional theory (TDDFT), to study nuclear structure and dynamics
in a unified way \cite{Simenel(review),Nakatsukasa(PTEP),Sky3D,Nakatsukasa(review)}.
Recently, we have applied the theory to study MNT and QF processes in various systems
at energies around the Coulomb barrier \cite{KS_KY_MNT,KS_KY_PNP,MyPhD,Bidyut(2015),
KS_KY_Ni-U,KS_SH_Kazimierz}. Applying particle-number projection (PNP) method \cite{Projection},
transfer cross sections were evaluated based on the TDHF theory. Comparisons with measured
cross sections revealed that the theory, being with no adjustable parameters, can describe
transfer cross sections quite well in accuracy comparable to the existing models \cite{KS_KY_MNT}. 
However, the calculated cross sections were of primary (excited) reaction products which are
to be deexcited through particle evaporation and/or fission. Because of this fact discrepancy
arises when compared with experimental data, especially for channels accompanying transfer
of many nucleons. The TDHF description of production cross sections has thus been beset with
the absence of deexcitation processes that limits its predictive power.

In this paper, a method, called TDHF+GEMINI, is proposed to cure the drawback of the
TDHF description. Namely, secondary deexcitation processes of primary reaction products,
both particle evaporation and fission, are simulated employing a state-of-the-art statistical model,
\fs{GEMINI}++ \cite{GEMINI++}. Those secondary processes are difficult to investigate within
TDHF, because of, \textit{e.g.}, its much longer timescales. The inputs of statistical-model calculations,
spin and excitation energy of primary reaction products, are extracted from the TDHF wavefunction
after collision, using an extended PNP method \cite{KS_KY_PNP}. The method is applied to
$^{40,48}$Ca+$^{124}$Sn, $^{40}$Ca+$^{208}$Pb, $^{58}$Ni+$^{208}$Pb,
$^{64}$Ni+$^{238}$U, and $^{136}$Xe+$^{198}$Pt reactions for which measured
cross sections are available. To demonstrate the accuracy of the proposed method is the
main purpose of this work.

It has been appreciated that TDHF provides valuable insight into complex many-body dynamics of
low-energy heavy ion reactions. However, it works with a single mean-field potential which is deterministically
associated with the initial condition. In other words, fluctuations in collective space is absent in the
TDHF dynamics. Indeed, it has been shown that the theory is optimized to describe the expectation
value of one-body observables \cite{BV(1981),Simenel(review)}, and fluctuations of them are known
to be severely underestimated \cite{Koonin(1977),Davies(1978),Dasso(1979),Simenel(2011)}.
How and to what extent beyond-mean-field fluctuations play a role in MNT reactions is an open question.
Moreover, inter-nucleon correlations are also not included in TDHF. Outcomes of MNT reactions may reflect
effects of inter-nucleon correlations, because of possible transfer of a correlated-pair or a cluster of nucleons.
Recently, it has become possible to pursue microscopic simulations of heavy ion reactions including the pairing
correlations \cite{Scamps(2012),Scamps(2013),Ebata(2014),Ebata(2015),Hashimoto(2016),MSW(2017),
SMW(2017),SWM(2017)}. It should be noted here that the proposed method can, in principle, be extended
to incorporate with the pairing correlations. In the present paper, however, we will focus on a treatment without
pairing and leave further extension/application to include the pairing correlations as a future task. Nevertheless,
it has to be noted here that, by extending the application of the theoretical framework based on the TDHF
theory as far as possible, this work will shed light on the validity of the theoretical framework that works
with a single mean-field potential, without inter-nucleon correlations.

\vspace{1mm}
The article is organized as follows.
In Sec.~\ref{Sec:method}, the methodology of TDHF+GEMINI is outlined.
In Sec.~\ref{Sec:results}, numerical results for various reactions are presented and are compared with available experimental data.
In Sec.~\ref{Sec:summary}, a summary of the present work is given.

\section{TDHF+GEMINI}{\label{Sec:method}}

In this Section, we focus on the analysis of the TDHF wavefunction
using the PNP method and its coupling with statistical-model calculations.
For details of the TDHF theory and its application to nuclear systems, see,
\textit{e.g.}, Refs.~\cite{Negele(review),Simenel(review),Nakatsukasa(PTEP),
Sky3D,Nakatsukasa(review)}, and references therein.

\subsection{Cross sections for primary products}

Let $A_\mu$, $Z_\mu$, and $N_\mu$, respectively, be mass, charge, and
neutron numbers of a projectile ($\mu={\rm P}$) and a target ($\mu={\rm T}$).
The total numbers of neutrons and protons in the system are $N^{(n)}=
N_{\rm P}+N_{\rm T}$ and $N^{(p)}=Z_{\rm P}+Z_{\rm T}$, respectively.
The total number of nucleons is denoted as $A=N^{(n)}+N^{(p)}$.

Suppose that we have performed a TDHF calculation for a reaction at a given
energy $E$ and an impact parameter $b$, and we observed generation of binary
reaction products. Now we have a many-body wavefunction at a certain time
$t=t_{\rm f}$ after collision, which is given by a single Slater determinant:
\begin{equation}
\Psi(\bs{r}_1\sigma_1q_1,\dots,\bs{r}_A\sigma_Aq_A,t_{\rm f})
= \frac{1}{\sqrt{A!}}\det\bigl\{\psi_i^{(q_j)}(\bs{r}_j\sigma_j,t_{\rm f})\bigr\},
\label{Eq:Psi_f}
\end{equation}
where $\psi_i^{(q)}(\bs{r}\sigma,t_{\rm f})$ is $i$th single-particle orbital at $t=t_{\rm f}$
with spatial, spin, and isospin coordinates, $\bs{r}$, $\sigma$, and $q$, respectively.
Our aim is to evaluate production cross sections for \textit{secondary} reaction products
based on the TDHF wavefunction after collision, Eq.~(\ref{Eq:Psi_f}).

Because of possible nucleon transfer processes, the TDHF wavefunction after
collision is, in general, not an eigenstate of a number operator in a subspace $V$
that contains one of the reaction products, but a superposition of states with
different particle-number distributions:
\begin{equation}
\big|\Psi\bigr> = \sum_{N,Z} \big|\Psi_{N,Z}\bigr>,
\end{equation}
where $N$ and $Z$ specify neutron and proton numbers of a reaction product
inside the spatial region $V$, respectively. Here and henceforth, brackets, such as
$\big|\Psi\bigr>$ and $\big|\psi_i^{(q)}\bigr>$, are often used omitting indexes
to simplify notations. $\big|\Psi_{N,Z}\bigr>$ can be expressed as
\begin{equation}
\big|\Psi_{N,Z}\bigr> = \hat{P}_N^{(n)} \hat{P}_Z^{(p)} \big|\Psi\bigr>,
\end{equation}
where $\hat{P}_n^{(q)}$ is the PNP operator for neutrons ($q=n$) or protons ($q=p$),
\begin{equation}
\hat{P}_n^{(q)} = \frac{1}{2\pi}\int_0^{2\pi}e^{\zI(n-\hat{N}_V^{(q)})\theta}d\theta.
\end{equation}
$\hat{N}_V^{(q)}$ is the number operator for neutrons ($q=n$) or protons ($q=p$)
in the spatial region $V$,
\begin{equation}
\hat{N}_V^{(q)} = \int_V \sum_{i=1}^{N^{(q)}} \delta(\bs{r}-\hat{\bs{r}}_i)\,d\bs{r} = \sum_{i=1}^{N^{(q)}} \Theta_V(\hat{\bs{r}}_i),
\end{equation}
where
\begin{equation}
\Theta_V(\bs{r}) = \left\{
\begin{array}{ccc}
1 & \;\mbox{for} & \bs{r} \in V,\\[0.5mm]
0 & \;\mbox{for} & \bs{r} \notin V.
\end{array}\right.
\end{equation}

The probability that a reaction product composed of $N$ neutrons
and $Z$ protons is produced, $P_{N,Z}$, is given by
\begin{equation}
P_{N,Z} = \bigl<\Psi_{N,Z}\big|\Psi_{N,Z}\bigr>
= P_N^{(n)} P_Z^{(p)}.
\label{Eq:P_NZ}
\end{equation}
Note that $P_{N,Z}$ is a product of probabilities for neutrons $P_N^{(n)}$
and for protons $P_Z^{(p)}$ in TDHF. These probabilities can be expressed
in terms of the single-particle orbitals as
\begin{equation}
P_n^{(q)} = \frac{1}{2\pi}\int_0^{2\pi}e^{\zI n\theta}\det\mathcal{B}^{(q)}(\theta)\,d\theta,
\end{equation}
where
\begin{eqnarray}
\bigl(\mathcal{B}^{(q)}(\theta)\bigr)_{ij}
&=& \sum_\sigma\int \psi_i^{(q)*}(\bs{r}\sigma)\psi_j^{(q)}(\bs{r}\sigma,\theta)\,d\bs{r} \0\\
&\equiv& \bigl<\psi_i^{(q)}\big|\psi_j^{(q)}(\theta)\bigr>
\end{eqnarray}
with
\begin{equation}
\psi_i^{(q)}(\bs{r}\sigma,\theta) = \bigl[ \Theta_{\bar{V}}({\bf r})+e^{-\zI\theta}\Theta_{V}({\bf r}) \bigr]
\psi_i^{(q)}(\bs{r}\sigma).
\end{equation}

By repeating TDHF calculations for various impact parameters $b$ at a given
incident energy $E$, we obtain $P_{N,Z}(b,E)$. The production cross section
for a \textit{primary} reaction product composed of $N$ neutrons and $Z$
protons before secondary deexcitation is then given by
\begin{equation}
\sigma_{N,Z}(E) = 2\pi \int_{b_{\min}}^{b_{\rm cut}} b\,P_{N,Z}(b,E)\,db,
\label{Eq:sigma_primary}
\end{equation}
where $b_{\rm min}$ is the minimum impact parameter for binary reactions,
inside which fusion reactions take place; $b_{\rm cut}$ is a cutoff impact
parameter for the numerical integration. Note that if $b_{\rm cut}$ is
chosen large enough it merely affects magnitude of the cross section for
the elastic scattering.

\subsection{Total angular momentum}\label{Sec:J}

In Ref.~\cite{KS_KY_PNP}, the PNP method has been extended to calculate
the expectation value of operators. The method allows to evaluate the total
angular momentum, $J$, of a reaction product in each transfer channel, which
is an input of a statistical-model calculation.

The idea \cite{KS_KY_PNP} was to introduce operators for a reaction product
inside the spatial region $V$. In the case of the total angular momentum operator,
it can be expressed as
\begin{equation}
\hat{\bs{J}}_V = \sum_{i=1}^A \Theta_V(\hat{\bs{r}}_i)\,\hat{\bs{j}}_i,
\end{equation}
where $\hat{\bs{j}}_i=(\hat{\bs{r}}_i-\bs{R}_{\rm c.m.})\times\hat{\bs{p}}_i
+\hat{\bs{s}}_i$. $\bs{R}_{\rm c.m.}$ is the center-of-mass position of the
reaction product, $\hat{\bs{p}}_i$ and $\hat{\bs{s}}_i$ are the momentum
and the spin operators, respectively,

The expectation value of the total angular momentum of a reaction product
composed of $N$ neutrons and $Z$ protons is then defined as
\begin{equation}
\bs{J}_{N,Z}
= \frac{\bigl<\Psi_{N,Z}\big|\hat{\bs{J}}_V\big|\Psi_{N,Z}\bigr>}{\bigl<\Psi_{N,Z}\big|\Psi_{N,Z}\bigr>}
= \bs{J}_N^{(n)} + \bs{J}_Z^{(p)}.
\label{Eq:J_NZ_PNP}
\end{equation}
It satisfies an identity, $\bigl<\Psi\big|\hat{\bs{J}}_V\big|\Psi\bigr>
=\sum_{N,Z}P_{N,Z}\bs{J}_{N,Z}$. The contribution from neutrons
($q=n$) or protons ($q=p$) is given by
\begin{equation}
\bs{J}_n^{(q)} = \frac{1}{2\pi P_n^{(q)}}\hspace{-1mm}\int_0^{2\pi}\hspace{-2mm}
e^{\zI n\theta} \det\mathcal{B}^{(q)}(\theta) \hspace{-0.6mm} \sum_{i=1}^{N^{(q)}}
\bigl<\psi_i^{(q)}\big|\hat{\bs{j}}\big|\tilde{\psi}_i^{(q)}(\theta)\bigr>_V d\theta,
\label{Eq:J_n_q}
\end{equation}
where
\begin{equation}
\tilde{\psi}_i^{(q)}(\bs{r}\sigma,\theta) \equiv
\sum_{j=1}^{N^{(q)}} \psi_j^{(q)}(\bs{r}\sigma,\theta) \bigl(\mathcal{B}^{(q)}(\theta)\bigr)_{ji}^{-1}.
\end{equation}
Note that $\{\tilde{\psi}_i(\theta)\}$ are biorthonormal to $\{\psi_i\}$,
\textit{i.e.}, $\bigl<\psi_i\big|\tilde{\psi}_j(\theta)\bigr>=\delta_{ij}$.
The subscript $V$ of the bracket in Eq.~(\ref{Eq:J_n_q}) indicates that the
spatial integration is taken only over the spatial region $V$. In practice, the
total angular momentum perpendicular to the reaction plane will be used as
an input for statistical-model calculations. It will be denoted simply as $J_{N,Z}$.

\subsection{Excitation energy}\label{Sec:Eex}

Applying the PNP method, we can also evaluate the excitation energy,
$E^*$, of a reaction product  in each transfer channel \cite{KS_KY_PNP}.
The energy expectation value of a reaction product composed of $N$ neutrons
and $Z$ protons is defined as
\begin{equation}
E_{N,Z} = \frac{\bigl<\Psi_{N,Z}\big|\hat{H}_V\big|\Psi_{N,Z}\bigr>}{\bigl<\Psi_{N,Z}\big|\Psi_{N,Z}\bigr>},
\label{Eq:E_NZ}
\end{equation}
where $\hat{H}_V$ is a Hamiltonian for a reaction product inside the
spatial region $V$. There also follows an identity, $\bigl<\Psi\big|\hat{H}_V
\big|\Psi\bigr>=\sum_{N,Z}P_{N,Z}E_{N,Z}$.

In practice, one may work with an energy density functional (EDF).
In such a case, $E_{N,Z}$ is given by
\begin{eqnarray}
E_{N,Z} &=& \frac{1}{4\pi^2P_N^{(n)}P_Z^{(p)}}\int_0^{2\pi}\int_0^{2\pi} e^{\zI(N\theta+Z\varphi)} \0\\[0.5mm]
&&\times \det\mathcal{B}(\theta,\varphi) \int_V \mathcal{E}(\bs{r},\theta,\varphi)\,d\bs{r}\,d\theta\,d\varphi,
\label{Eq:E_NZ_PNP}
\end{eqnarray}
where $\det\mathcal{B}(\theta,\varphi)=\det\mathcal{B}^{(n)}(\theta)
\det\mathcal{B}^{(p)}(\varphi)$. $\mathcal{E}(\bs{r},\theta,\varphi)$
denotes an EDF kernel which has the same form as the EDF used, but is composed
of complex mixed densities, \textit{e.g.}, $\rho_q(\bs{r},\theta)=\sum_{i,\sigma}
\psi_i^{(q)*}(\bs{r}\sigma)\tilde{\psi}_i^{(q)}(\bs{r}\sigma,\theta)$, etc.

For our purpose, Eq.~(\ref{Eq:E_NZ}) or Eq.~(\ref{Eq:E_NZ_PNP}) should be evaluated
in the rest frame of the reaction product in order to remove kinetic energy associated
with center-of-mass translational motion. Regarding it as internal energy of the reaction
product, excitation energy can be evaluated as
\begin{eqnarray}
E^*_{N,Z} = E_{N,Z} - E_{N,Z}^{\rm g.s.},
\label{Eq:Eex}
\end{eqnarray}
where $E_{N,Z}^{\rm g.s.}$ denotes energy of a nucleus specified by $N$ neutrons
and $Z$ protons in its Hartree-Fock ground state.

\subsection{Cross sections for secondary products}

Having $P_{N,Z}$, $J_{N,Z}$, and $E_{N,Z}^*$ at hand, we can apply a statistical
model to obtain production cross sections for \textit{secondary} reaction products.
In this paper, \fs{GEMINI}++ \cite{GEMINI++} is employed. For a given set of
($N$,~$Z$,~$E^*$,~$J$), \fs{GEMINI}++ provides a sequence of statistically-selected
binary decays, including both particle evaporation and fission, until it becomes energetically
forbidden or improbable due to competing $\gamma$-ray emission.

Because of its statistical nature, a different sequence of binary decays can be obtained
with the same set of ($N$,~$Z$,~$E^*$,~$J$). To evaluate decay probabilities, one may
repeat computations of a binary-decay sequence, let's say, $N_{\rm trial}$ times.
Among the obtained $N_{\rm trial}$ decay sequences, one can count the number of
processes in which a sequence of deexcitation processes of a primary reaction product
composed of $N$ neutrons and $Z$ protons ends up with a nucleus specified by $N'$
neutrons and $Z'$ protons. Then, denoting it as $N_{N',Z'}$, the decay probability for
the process ($N,Z$)\,$\rightarrow$\,($N',Z'$) can be defined by
\begin{equation}
P_{\rm decay}(E_{N,Z}^*,J_{N,Z},N,Z; N',Z') = \frac{N_{N',Z'}}{N_{\rm trial}}.
\label{Eq:P_decay}
\end{equation}
The number $N_{\rm trial}$ sets a lower limit on the decay probabilities, since processes
with $P_{\rm decay}$\,$\lesssim 1/N_{\rm trial}$ will not be taken into account.

The cross section for a \textit{secondary} reaction product after evaporation and/or
fission processes is then evaluated as
\begin{eqnarray}
\tilde{\sigma}_{N',Z'}(E)
&=& 2\pi \int_{b_{\rm min}}^{b_{\rm cut}} b\,\widetilde{P}_{N',Z'}(b,E)\,db,
\label{Eq:sigma_secondary}
\end{eqnarray}
where $\widetilde{P}_{N',Z'}$ denotes the probability that a reaction product composed
of $N'$ neutrons and $Z'$ protons is produced after secondary deexcitation processes:
\begin{eqnarray}
\widetilde{P}_{N',Z'} = \hspace{-1mm}\sum_{N \ge N'}\hspace{-0.5mm}\sum_{Z \ge Z'}\hspace{-0.8mm}
P_{N,Z} P_{\rm decay}(E_{N,Z}^*,J_{N,Z},N,Z; N'\hspace{-1mm},Z').\0\\[-3mm]
\label{Eq:P_NZ_secondary}
\end{eqnarray}
Note that the input quantities, $P_{N,Z}$, $J_{N,Z}$, and $E_{N,Z}^*$,
are dependent on the incident energy $E$ and the impact parameter $b$.

\subsection{Computational details}

For the TDHF calculation and the PNP analysis, our own computational code has
been extended and applied \cite{MyPhD}. In the code, single-particle orbitals are
represented on three-dimensional Cartesian coordinates (without any symmetry
restrictions) with isolated boundary conditions. The lattice spacing is set to 0.8~fm.
Spatial derivatives are computed with 11-point finite-difference formula. For time
evolution, 4th-order Taylor expansion method is used with a single predictor-corrector
step. The time step is set to 0.2~fm/$c$. Coulomb potential is computed using Fourier
transformations. For the EDF, Skyrme SLy5 functional \cite{Chabanat} is employed.

Note that results of the PNP analysis with Eqs.~(\ref{Eq:P_NZ}), (\ref{Eq:J_NZ_PNP}),
and (\ref{Eq:E_NZ}) are invariant under unitary transformations for the single-particle
orbitals, $\psi_i'(\bs{r}\sigma)=\sum_j\mathcal{U}_{ij}\psi_j(\bs{r}\sigma)$, where
$\mathcal{U}$ is a unitary matrix: it only changes the phase of the many-body wavefunction,
\textit{i.e.}, $\Psi'=\det\mathcal{U}\,\Psi$. Thus, the quantities are well-defined and the
results may merely depend on the accuracy of the numerical integration and the choice of
the spatial region $V$. The latter is expected to be small in the case where reaction products
are well separated spatially. In the present analysis, $V$ is taken as a sphere with a radius
$R_V$ around the center-of-mass of a reaction product. Indeed, it has been confirmed, by
varying $R_V$ from 11~fm to 15~fm for the $^{40}$Ca+$^{124}$Sn system, that $J_{N,Z}$
and $E_{N,Z}$ of lighter fragments are affected only less than $0.01\,\hbar$ and a few tens
of keV, respectively.

The numerical integration over the gauge angle $[0,2\pi]$ is performed using the trapezoidal
rule with an $M$-point uniform mesh. $P_{N,Z}$ and $J_{N,Z}$ can be stably computed for
all transfer channels and do not depend on $M$, if it is taken larger than $M\approx100$--200
for the systems analyzed in the present paper. The accuracy of $E_{N,Z}$, on the other hand,
is somewhat worse, which depends on the probability $P_{N,Z}$. For example, for different values
of $M$ (200--500), $E_{N,Z}$ varies roughly several keV for a main process with $P_{N,Z}\approx
10^{-1}$, whereas for a process with smaller probabilities, $P_{N,Z} \approx10^{-5}$, difference
can be several hundreds of keV. Moreover, absolute value of $E_{N,Z}$ becomes unphysically large
for processes with tiny probabilities, $P_{N,Z}\lesssim10^{-5}$, as was observed for a lighter system
\cite{KS_KY_PNP}. In the present paper, $M=300$ and $R_V=15$~fm are utilized. It ensures
at least 1-MeV accuracy of $E_{N,Z}$, which would be sufficient for the present purpose of
quantifying effects of secondary deexcitation processes on production cross sections.

The ground-state energy of nuclei, $E_{N,Z}^{\rm g.s.}$ in Eq.~(\ref{Eq:Eex}), is
computed employing a cubic box with $20.8$~fm on each side. The mesh spacing is
set to 0.8~fm, as in the TDHF calculations. To find the energy minimum solution, avoiding
those in local minima, static Hartree-Fock calculations are first performed with constraints on
various deformation parameters ($\beta=0$, and $\beta=0.1,0.2$ with $\gamma=0^\circ,
30^\circ, 60^\circ$). After getting a moderately convergent solution, the constraints on
$\beta$ and $\gamma$ are released, and the energy is re-minimized. The state with the
lowest energy among the seven candidates is regarded as the Hartree-Fock ground state.
The ground-state energies of various isotopes with $Z\le 35$ are prepared for the evaluation
of excitation energies of respective transfer products. Equal-filling treatment is adopted for
odd number of nucleons, where the last nucleon resides in a time-reversal pair of orbitals
with half occupation.

To describe secondary processes, the statistical compound-nucleus deexcitation model,
\fs{GEMINI}++ \cite{GEMINI++}, is employed. \fs{GEMINI}++ is an improved version of
a statistical model, \fs{GEMINI}, developed by R.J. Charity \cite{GEMINI}. It takes into
account not only evaporation of light particles, \textit{i.e.}, $n$, $p$, $d$, $t$, $^{3}$He,
$\alpha$, $^{6}$He, $^{6-8}$Li, and $^{7-10}$Be, with the Hauser-Feshbach formalism
\cite{Hauser-Feshbach}, but also emission of heavier fragments with a binary-decay formalism
of L.G.~Moretto \cite{Moretto1975}, as well as fission based on the Bohr-Wheeler formalism
\cite{Bohr-Wheeler}. The default setting of the code is used for all reactions analyzed in the
present paper. The ingredients of the statistical model have been parametrized and determined
so as to allow a good systematic description of the evaporation spectra for the entire mass region.
Detailed discussions on various modifications and fine-tuning of the model parameters that were
implemented in the \fs{GEMINI}++ code can be found in Refs.~\cite{Charity2010,Mancusi2010}.
It has been tested, for all the systems under study (see Table~\ref{table}), that $N_{\rm trial}=100$,
1000, and 10000 provide almost identical cross sections. (Even $N_{\rm trial}=10$ gives very
similar cross sections, because of neutron evaporation dominance in those reactions.)
In the following, results obtained with $N_{\rm trial}=1000$ are presented.

\section{RESULTS AND DISCUSSION}{\label{Sec:results}}

In the present paper, TDHF+GEMINI is applied to six reaction systems listed in Table~\ref{table},
for which precise experimental data are available. One should note that the reactions investigated
cover a range of systems which are expected to show different properties of the reaction dynamics.
In reactions of projectile and target nuclei with different $N/Z$ ratios, the charge equilibration process
takes place \cite{Freiesleben(1984)}, where neutrons and protons are transferred toward the opposite
directions, to reduce the $N/Z$ imbalance of the colliding nuclei. Fusion reaction in a system with a large
charge product exceeding a critical value, $Z_{\rm P}Z_{\rm T}\gtrsim1600$, is known to be substantially
hindered \cite{Sahm(1984)}. Those characteristic quantities are also listed in Table~\ref{table}.

A detailed analysis of the TDHF results for MNT processes in the $^{40,48}$Ca+$^{124}$Sn,
$^{40}$Ca+$^{208}$Pb, and $^{58}$Ni+$^{208}$Pb reactions was carried out in Ref.~\cite{KS_KY_MNT}.
Recently, TDHF calculations for MNT and QF processes in the $^{64}$Ni+$^{238}$U reaction were
also performed. Comprehensive discussions on interplay between orientations of deformed $^{238}$U
and quantum shells in the reaction dynamics can be found in Ref.~\cite{KS_KY_Ni-U}. Here, the same
TDHF wavefunctions are re-analyzed applying TDHF+GEMINI.

TDHF calculations for the $^{136}$Xe+$^{198}$Pt reaction have been newly performed.
The ground state of $^{136}$Xe and $^{198}$Pt turned out to be slightly deformed in a
triaxial shape with deformation parameters, ($\beta\simeq0.06$, $\gamma\simeq29^\circ$)
and ($\beta\simeq0.12$, $\gamma\simeq33^\circ$), respectively. Those nuclei were placed
in such a way that the axis around which $\bigl|Q_{22}\bigr|$ takes the smallest value is set
perpendicular to the reaction plane. TDHF calculations were performed for various impact
parameters in a range of $[0,\,12]$ (fm). No fusion reaction was observed for all impact
parameters. Reaction mechanisms including incident energy dependence will be investigated
in the forthcoming paper. In this paper, we will focus on the production cross sections for the
$E_{\rm c.m.}\simeq644.98$~MeV case.

\begin{table}[t]
\caption{
A list of information of the reactions investigated.
}
\label{table}
\begin{center}
\begin{tabular}{cccccc}
\hline\hline
System & $E_{\rm c.m.}$ (MeV) & $N_{\rm P}/Z_{\rm P}$ & $N_{\rm T}/Z_{\rm T}$ & $Z_{\rm P}Z_{\rm T}$ & Expt. \\
\hline
$^{40}$Ca+$^{124}$Sn  & 128.54 & 1.00 & 1.48 & 1000 & \cite{Corradi(40Ca+124Sn)} \\
$^{48}$Ca+$^{124}$Sn  & 125.44 & 1.40 & 1.48 & 1000 & \cite{Corradi(48Ca+124Sn)} \\
$^{40}$Ca+$^{208}$Pb  & 208.84 & 1.00 & 1.54 & 1640 & \cite{Szilner(40Ca+208Pb)} \\
$^{58}$Ni+$^{208}$Pb   & 256.79 & 1.07 & 1.54 & 2296 & \cite{Corradi(58Ni+208Pb)} \\
$^{64}$Ni+$^{238}$U     & 307.35 & 1.29 & 1.57 & 2576 & \cite{Corradi(64Ni+238U)} \\
$^{136}$Xe+$^{198}$Pt & 644.98 & 1.52 & 1.54 & 4212 & \cite{Watanabe(136Xe+198Pt)} \\
\hline\hline
\end{tabular}\vspace{-2mm}
\end{center}
\end{table}

It would be useful to first digest success and failure of the TDHF description. Because of the microscopic
nature of the TDHF theory, it was not known, before our study \cite{KS_KY_MNT}, that to what extent
TDHF can quantitatively describe cross sections for MNT processes. By performing a number of TDHF
calculations for various impact parameters $b$ and by plugging $P_{N,Z}(b)$ into Eq.~(\ref{Eq:sigma_primary}),
one can evaluate production cross sections for primary reaction products, based on the TDHF calculations.
It should be noted here that no empirical parameters are introduced that can be adjusted to reproduce
experimental data.

\begin{figure}[t]
\begin{center}
\includegraphics[width=8.6cm]{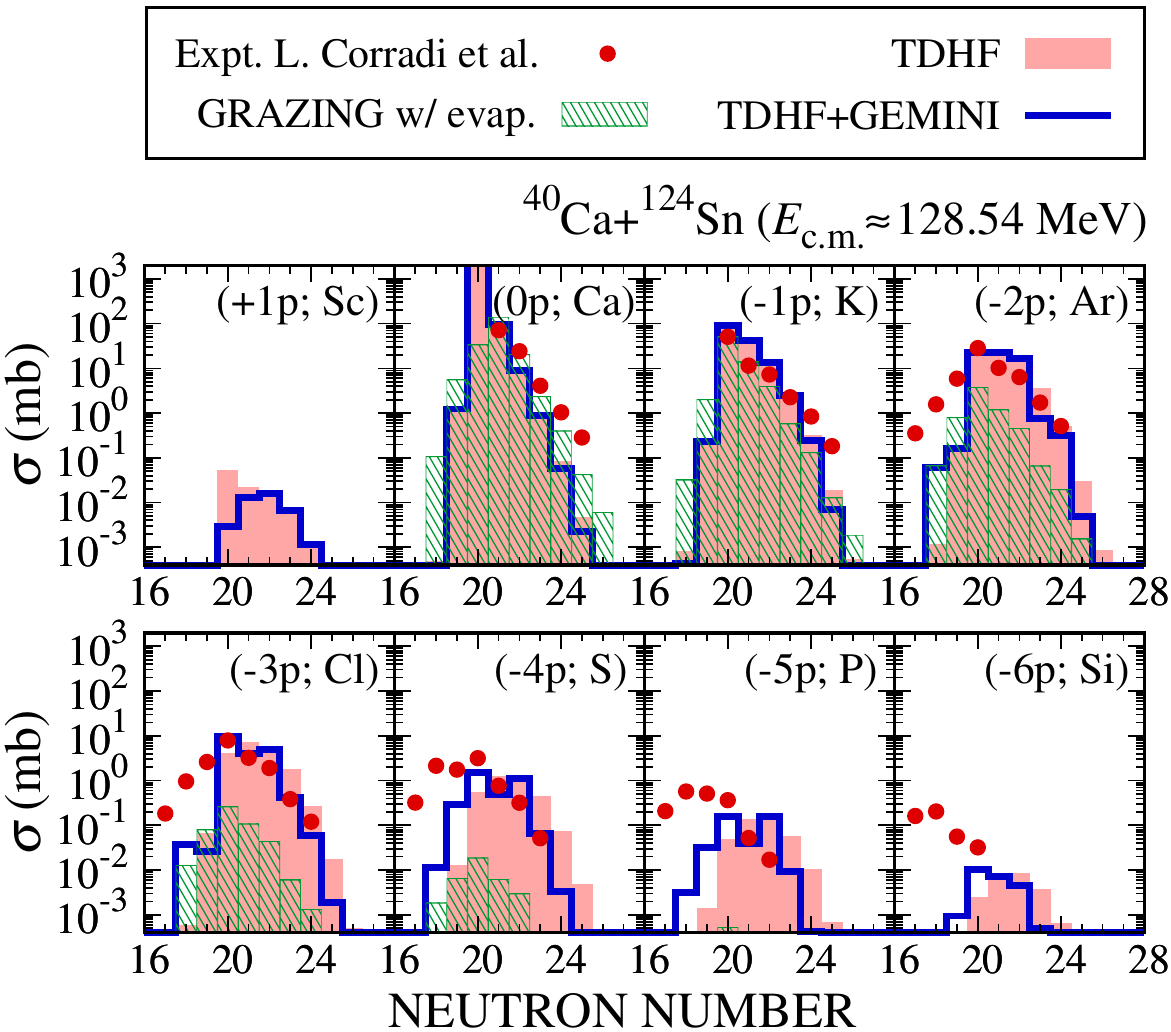}\vspace{-1mm}
\caption{
   Production cross sections for lighter fragments in the $^{40}$Ca+$^{124}$Sn reaction
   at $E_{\rm c.m.}\simeq128.54$~MeV. Each panel shows cross sections for different
   proton-transfer channel. The horizontal axis is the neutron number of the fragments.
   Red filled circles denote the experimental data \cite{Corradi(40Ca+124Sn)}. Red filled
   areas represent TDHF results for primary reaction products. Cross sections for secondary
   reaction products obtained with TDHF+GEMINI are shown by blue solid lines. For comparison,
   \fs{GRAZING} results \cite{GRAZING-online} are also shown by green shaded histograms.
}\vspace{-7.5mm}
\label{FIG:NTCS_40Ca+124Sn}
\end{center}
\end{figure}

\begin{figure}[b]
\begin{center}\vspace{-2.5mm}
\includegraphics[width=8.6cm]{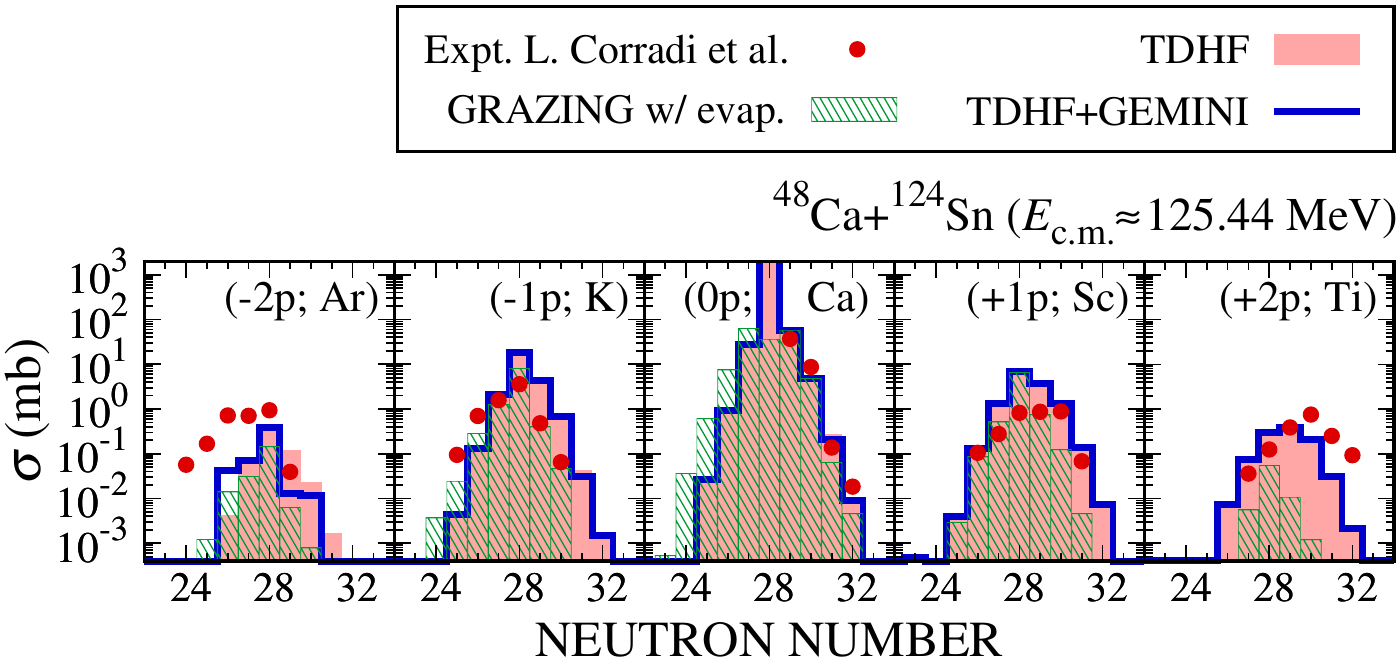}\vspace{-1mm}
\caption{
   Same as Fig.~\ref{FIG:NTCS_40Ca+124Sn}, but for the $^{48}$Ca+$^{124}$Sn
   reaction at $E_{\rm c.m.}\simeq125.44$~MeV. The experimental data were reported
   in Ref.~\cite{Corradi(48Ca+124Sn)}.
}
\label{FIG:NTCS_48Ca+124Sn}
\end{center}
\end{figure}

\begin{figure}[t]
\begin{center}
\includegraphics[width=8.6cm]{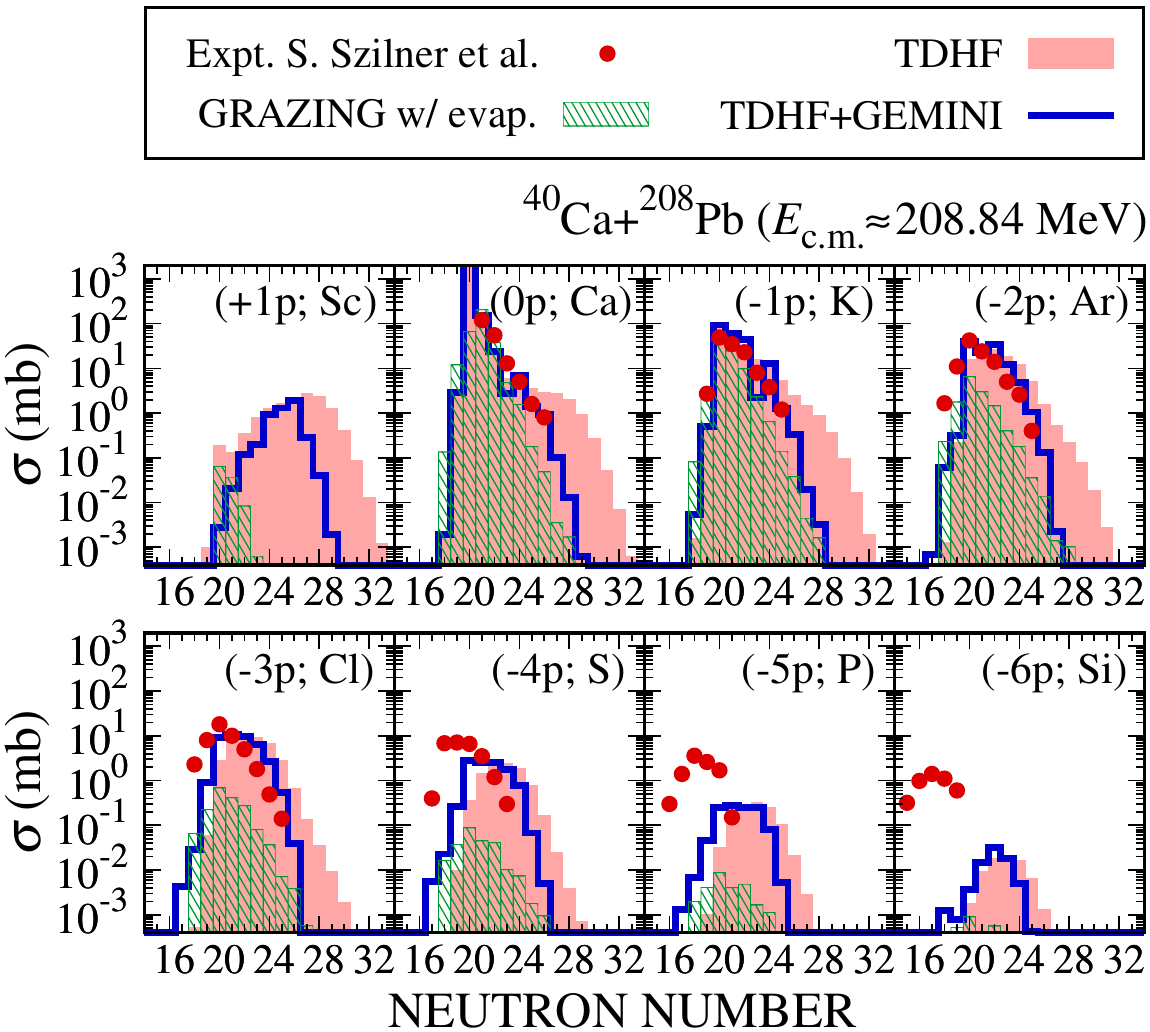}\vspace{-1mm}
\caption{
   Same as Figs.~\ref{FIG:NTCS_40Ca+124Sn} and \ref{FIG:NTCS_48Ca+124Sn},
   but for the $^{40}$Ca+$^{208}$Pb reaction at $E_{\rm c.m.}\simeq208.84$~MeV.
   The data for $E_{\rm c.m.}\simeq197.10$~MeV were also analyzed, getting very similar
   results (not shown). The experimental data were reported in Ref.~\cite{Szilner(40Ca+208Pb)}.
}\vspace{-7.5mm}
\label{FIG:NTCS_40Ca+208Pb_E249}
\end{center}
\end{figure}

\begin{figure}[b]
\begin{center}\vspace{-2.5mm}
\includegraphics[width=8.6cm]{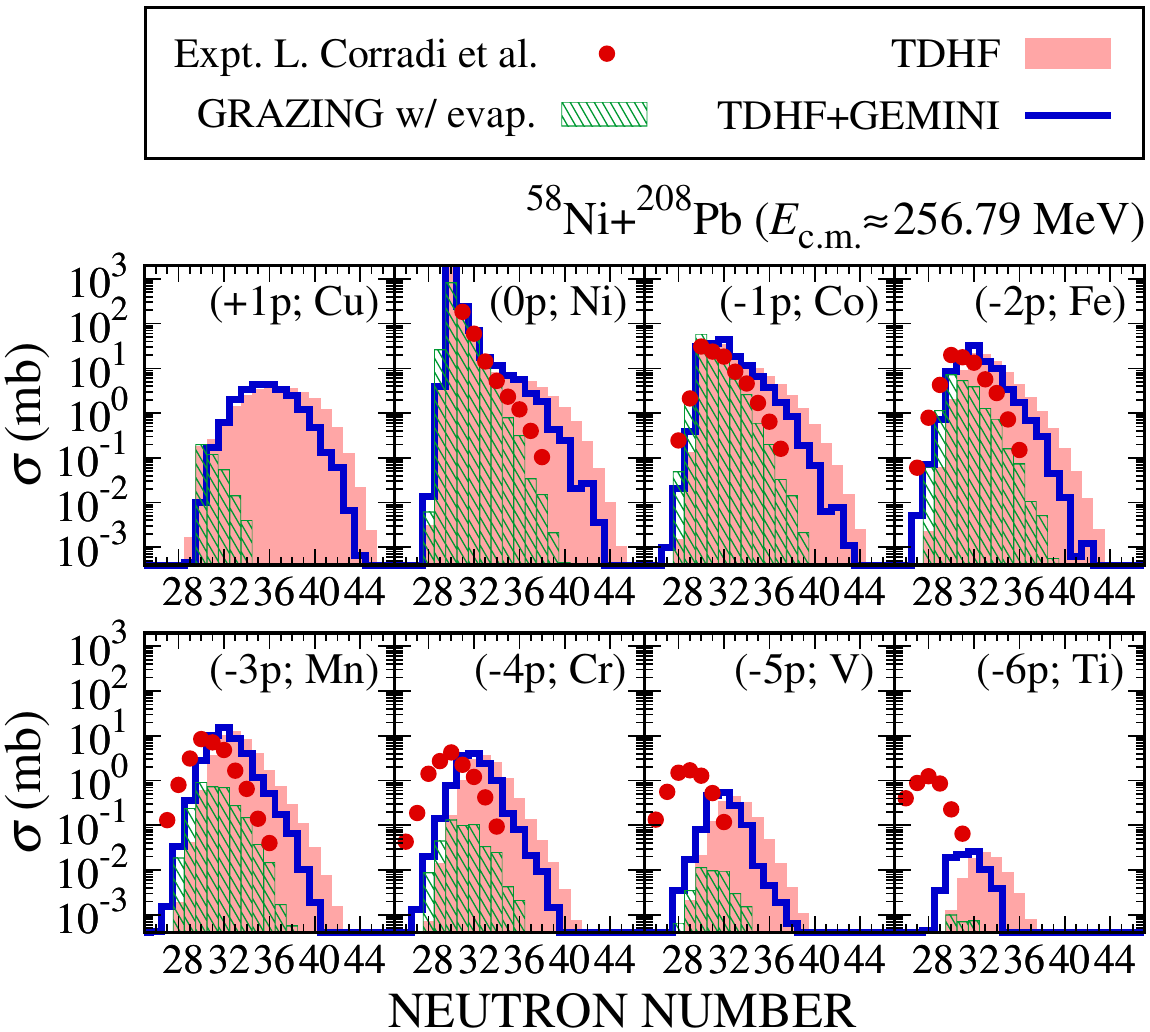}\vspace{-1mm}
\caption{
   Same as Figs.~\ref{FIG:NTCS_40Ca+124Sn}--\ref{FIG:NTCS_40Ca+208Pb_E249},
   but for the $^{58}$Ni+$^{208}$Pb reaction at $E_{\rm c.m.}\simeq256.79$~MeV.
   The experimental data were reported in Ref.~\cite{Corradi(58Ni+208Pb)}.
}
\label{FIG:NTCS_58Ni+208Pb}
\end{center}
\end{figure}

The calculations were first performed for the $^{40,48}$Ca+$^{124}$Sn, $^{40}$Ca+$^{208}$Pb, and
$^{58}$Ni+$^{208}$Pb reactions \cite{KS_KY_MNT}. In Figs.~\ref{FIG:NTCS_40Ca+124Sn}--\ref{FIG:NTCS_58Ni+208Pb},
production cross sections for lighter (projectile-like) fragments are shown. Cross sections are classified
according to the number of transferred protons~$x$, indicated by ($\pm x$p;~X), where X stands for
the corresponding element. The plus sign corresponds to proton transfer from the target to the projectile
(pickup), while the minus sign corresponds to the opposite (stripping). The horizontal axis is the neutron
number of the fragments. Experimental data are shown by red filled circles. The cross sections for \textit{primary}
reaction products obtained from the TDHF calculations \cite{KS_KY_MNT} are shown by red filled areas.

It is evident from the figures that TDHF nicely captures the main features of the reaction dynamics.
In the reactions with a large $N/Z$ asymmetry ($^{40}$Ca+$^{124}$Sn, $^{40}$Ca+$^{208}$Pb,
and $^{58}$Ni+$^{208}$Pb), neutron-pickup and proton-stripping are favored, because of the
charge equilibration process (Figs.~\ref{FIG:NTCS_40Ca+124Sn},\,\ref{FIG:NTCS_40Ca+208Pb_E249},\,\ref{FIG:NTCS_58Ni+208Pb}).
On the other hand, in the $^{48}$Ca+$^{124}$Sn reaction where $N/Z$ ratios of the projectile and
the target are already very close to each other, the number of transferred nucleons becomes very small
on average, resulting in transfer of neutrons and protons in both directions (Fig.~\ref{FIG:NTCS_48Ca+124Sn}).

From a careful look at the figures, one can see that TDHF quantitatively reproduces the measured
cross sections for few-nucleon transfer processes: \textit{e.g.} processes with pickup of a few
neutrons in ($0p$), ($\pm1p$), and ($-2p$) proton-transfer channels. The agreements are noteworthy,
since no adjustable parameters are included in the calculations. However, there exist discrepancies
between the TDHF results and the experimental data. Namely, the measurements show that substantial
cross sections for fragments with a smaller number of neutrons emerge when more than one protons are
transferred. TDHF fails to reproduce the observed tendency. Moreover, cross sections for multi-proton
stripping processes are considerably underestimated by the TDHF calculations. On the other hand,
in the $^{40}$Ca+$^{208}$Pb and $^{58}$Ni+$^{208}$Pb reactions, TDHF overestimates cross
sections for pickup of many neutrons in ($0p$), ($-1p$), and ($-2p$) channels. The latter originated
from trajectories at small impact parameters accompanying large energy losses, where dynamics of
a thick-neck formation and its breaking is responsible for the amount of nucleon transfer (see Ref.~\cite{KS_KY_MNT}
for a detailed discussion).

One can expect that the discrepancies may be remedied by including secondary deexcitation
processes in the TDHF description. After transfer of many nucleons reaction products can be
highly excited, and what were measured experimentally must be the cross sections for secondary
products after deexcitation processes. For example, if neutrons were evaporated from primary
reaction products, the resulting cross sections would shift toward less neutron-number side,
the left direction in the figures; if protons were emitted to the continuum, then the cross sections
would be re-distributed to channels classified as a process with a larger number of stripping of protons.
It is now feasible to disentangle the possible origins of the discrepancies owing to TDHF+GEMINI.

In Figs.~\ref{FIG:NTCS_40Ca+124Sn}--\ref{FIG:NTCS_58Ni+208Pb}, the cross sections for
\textit{secondary} reaction products obtained with TDHF+GEMINI are represented by blue solid
lines. In the evaluation of the cross sections, the PNP calculations for $J_{N,Z}$ and $E_{N,Z}^*$
presented in Secs.~\ref{Sec:J} and \ref{Sec:Eex}, respectively, were fully performed for all impact
parameters investigated. For comparison, online \fs{GRAZING} calculations \cite{GRAZING-online}
which include neutron-evaporation effects were also performed, and the results are shown by
green shaded histograms.

From the figures, one can see that the inclusion of secondary deexcitation processes improves
the description. For instance, TDHF+GEMINI quantitatively reproduces the experimental data,
mainly for neutron-pickup processes, accompanying stripping of several protons. The description of
the overestimated cross sections of neutron-pickup processes for ($0p$), ($-1p$), and ($-2p$)
channels in the $^{40}$Ca+$^{208}$Pb and $^{58}$Ni+$^{208}$Pb reactions is also improved.
However, the effects of secondary deexcitation processes are not large enough to remedy all the
discrepancies between the TDHF results and the experimental data. The peak positions of the
cross sections still locate at the larger neutron-number side compared to the experimental data.
Note that the magnitude of evaporation effects is very similar to the one observed in the
\fs{GRAZING} calculations as well as our prior attempts \cite{KS_KY_FUSION14,KS_KY_Maruhn}
with a simple evaporation model developed by I.~Dostrovsky \textit{et al.} \cite{Dostrovsky(1959)}.

It is worth emphasizing here that, although the peak positions are different compared to the
experimental cross sections, its absolute value is well reproduced by TDHF+GEMINI up to around
($-4p$) channel. The widely-used \fs{GRAZING} calculations provide few orders of magnitude smaller
cross sections, when more than one protons are transferred. In Refs.~\cite{Szilner(40Ca+208Pb),
Corradi(58Ni+208Pb)}, the $^{40}$Ca+$^{208}$Pb and $^{58}$Ni+$^{208}$Pb reactions
were analyzed by the CWKB semi-classical model. It provides quantitatively very similar results
as the \fs{GRAZING} calculations, \textit{i.e.}, the absolute values of the cross sections for
($-xp$) ($x\ge2$) channels were substantially underestimated. The authors of Refs.~\cite{Szilner(40Ca+208Pb),
Corradi(58Ni+208Pb)} phenomenologically introduced pair-transfer modes to explain the experimental
data. Although they could obtain better description of the cross sections by adjusting an additional
macroscopic form factor to reproduce the experimental cross section for pure two-proton stripping
process without neutron transfer, \textit{i.e.}, ($0n,-2p$) channel, the validity of the hypothetical
pair-transfer modes was unclear. The present results indicate that the pair-transfer processes may
play a minor role, at least for the above-barrier MNT processes, as the absolute value of the
cross sections for ($-2p$) channel is well reproduced without pairing by TDHF+GEMINI.

The remained discrepancies between the results of TDHF+GEMINI and the experimental
data would rather suggest a limit of the theoretical framework. For example, although we
could obtain cross sections for various transfer channels using the PNP method, they merely
come from distribution around the average trajectory described by a single Slater determinant.
Because of this fact, the peak positions are strongly correlated with the average number of
transfered nucleons. Since we do not have a mean-field potential for, \textit{e.g.}, ($-6p$)
channel, the underestimation of multi-proton stripping processes may be an artifact of the
usage of the single mean-field potential. In reality, the potential should be transfer-channel
dependent. When many protons are removed from the projectile, for instance, the potential
felt by neutrons inside the proton-removed (proton-added) nucleus would become shallower
(deeper) that may suppress/enhance neutron pickup/stripping processes. This transfer-channel
dependent potential may explain the observed discrepancies. Part of the effects may be seen
as beyond-mean-field fluctuations and correlations. One should note, however, that possible
underestimation of evaporation effects has not been excluded yet. For instance, $E^*_{N,Z}$
and $J_{N,Z}$ that were used for the statistical-model calculations are the expectation values
of energy and angular momentum of a reaction product in each transfer channel and, thus,
are averaged over all possible quantal states populated by the reaction. However, these
quantities should have certain distributions, and substantial part of evaporation processes
might originate from components which are not in the vicinity of the mean values. Those
distributions may not adequately be described within the TDHF theory, since for evaluation
of the variance of them proper description of two- and four-body observables is necessary.
In addition, \textit{e.g.}, the neglected prompt (pre-equilibrium) nucleon emissions before
compound-nucleus formation and any underestimation of negative $Q$-value effect would
enhance the particle evaporation. In any case, the present results by TDHF+GEMINI suggest
that, in order to fully reproduce experimental cross sections for MNT processes, one should
extend the theoretical framework that goes beyond the TDHF theory: \textit{e.g.} a variational
method of Balian and V\'en\'eroni \cite{BV(1981),Simenel(2011)}, stochastic mean-field
theory (SMF) \cite{Ayik(2008),Tanimura(2017)}, time-dependent density matrix theory (TDDM)
\cite{Shun-jin(1985),Tohyama(1985),Gong(1990),Tohyama(2016)}, time-dependent generator
coordinate method (TDGCM) \cite{Berger(1984),Simenel(2014)}, multiconfiguration TDHF
(MCTDHF) \cite{Kato(2004),Caillat(2005)}, and so forth.

It should be emphasized here that, although there still remain certain discrepancies as
discussed above, TDHF+GEMINI will be a powerful tool, especially for systems for which
no experimental data exist, to figure out optimal conditions to produce objective neutron-rich
isotopes, as a microscopic model without empirical parameters that has better accuracy
than the widely-used semi-classical models. In the following, let us thus consider further
applications of TDHF+GEMINI.

Although it is possible to perform the PNP calculations for $J_{N,Z}$ and $E_{N,Z}^*$ as
shown above, it requires quite large computational effort. For reactions involving heavy nuclei,
for instance, the number of discreet points for the numerical integration over the gauge angles
$\theta$ and $\varphi$ is typically a few hundreds. Then, to evaluate Eqs.~(\ref{Eq:J_n_q})
and (\ref{Eq:E_NZ_PNP}), one needs to compute derivatives of 3D complex spatial functions
many times, which could be as time-consuming as a TDHF calculation. Moreover, to compute
the excitation energies with Eq.~(\ref{Eq:Eex}), one has to compute the energy of nuclei in
their Hartree-Fock ground state in a wide mass region. It would also be problematic if one wants
to change the working EDF, then the ground-state energies have to be re-computed for the
evaluation of $E_{N,Z}^*$. One may consider to use experimental masses for ground-state
energies, $E_{N,Z}^{\rm g.s.}$, however, then the result would be dependent on quality
of the EDF itself.

Since it is not desirable to pay much effort for the computation of the input parameters of the
statistical-model calculations, let us consider a simpler evaluation. Namely, one may replace
$E_{N,Z}^*$ and $J_{N,Z}$ in Eq.~(\ref{Eq:P_decay}) with average quantities that can be
easily obtained from the TDHF wavefunction after collision. At the time $t=t_{\rm f}$ after collision,
one can compute the average total angular momentum as
\begin{equation}
\bar{J} = \bigl<\Psi\big|\hat{J}_{V}\big|\Psi\bigr> = \sum_{i=1}^A\bigl<\psi_i\big|\hat{j}\big|\psi_i\bigr>_V,
\label{Eq:Jave}
\end{equation}
where operators $\hat{J}_V$ and $\hat{j}$ are those for the component perpendicular
to the reaction plane. One can also compute such quantities as average mass and charge
numbers, $A_i$ and $Z_i$ ($i=1,2$), the relative vector connecting the center-of-mass
positions of the reaction products, $\bs{R}$, and its time derivative, $\dot{\bs{R}}$.
Then, the average total kinetic energy (TKE) of the outgoing fragments reads:
\begin{equation}
\mathrm{TKE} = \frac{1}{2}\mu\dot{\bs{R}}^2 + \frac{Z_1Z_2e^2}{|\bs{R}|},
\end{equation}
where $\mu=m_n A_1A_2/(A_1+A_2)$ is the reduced mass, being $m_n$ the nucleon
mass. From this quantity, average total excitation energy shared by the two fragments
may be evaluated as
\begin{equation}
\bar{E}_{\rm tot}^* = E - {\rm TKE} - Q,
\end{equation}
where $E$ is the incident relative energy and $Q$ denotes the $Q$ value of the reaction.
By putting the actual $Q$ value for each transfer channel, $E_{\rm tot}^*$ becomes effectively
transfer-channel dependent. To evaluate the $Q$ value for each transfer channel, experimental
masses from the latest atomic mass evaluation, A\fs{ME}2016 \cite{AME2016(I),AME2016(II)},
are utilized, whenever available; for nuclei whose mass has not been measured experimentally,
theoretical values from the newest version of the finite-range droplet model, FRDM(2012) \cite{FRDM(2012)},
are adopted. One may distribute the total excitation energy to respective reaction products
in such a way that it is proportional to their mass:
\begin{equation}
\bar{E}_{N,Z}^* = \frac{N+Z}{A_1+A_2}\,\bar{E}_{\rm tot}^*.
\label{Eq:Eex_TKEL}
\end{equation}
It is equivalent to assume that the thermal equilibrium is realized before formation
of the fragments. The assumption may not always be correct especially in a transitional
regime from quasielastic to deep-inelastic reactions. However, since in such a regime
energy loss is not that large and secondary processes play less important role, genuineness
of the assumption may not be significant.

\begin{figure}[t]
\begin{center}
\includegraphics[width=8.6cm]{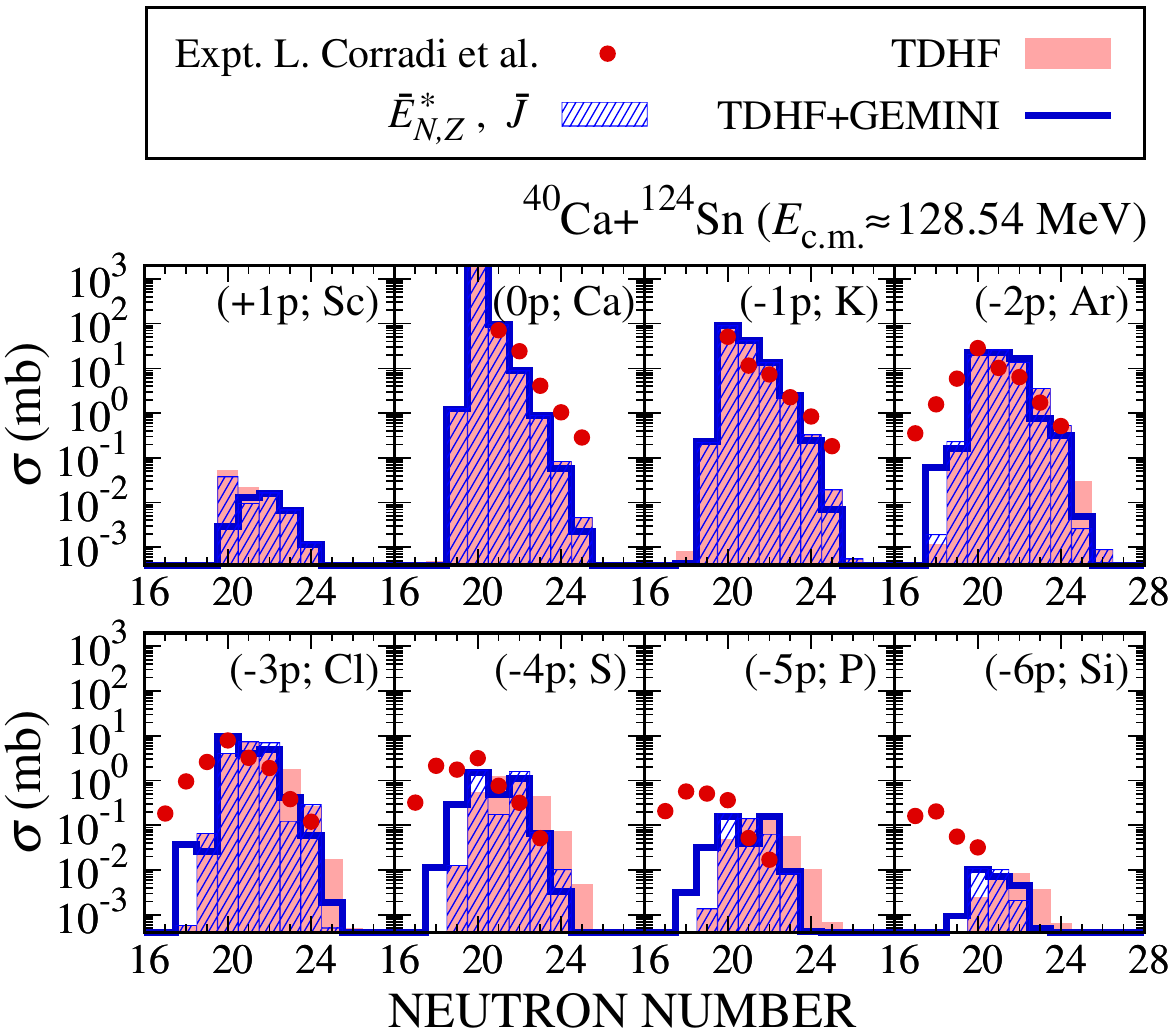}\vspace{-1mm}
\caption{
   Same as Fig.~\ref{FIG:NTCS_40Ca+124Sn} for the $^{40}$Ca+$^{124}$Sn
   reaction at $E_{\rm c.m.}\simeq128.54$~MeV, but cross sections by TDHF+GEMINI
   with a simpler treatment of $E^*$ and $J$ (see text for details) are shown by blue
   shaded histograms, instead of GRAZING results.
}\vspace{-7.5mm}
\label{FIG:NTCS_40Ca+124Sn_ave}
\end{center}
\end{figure}

\begin{figure}[b]
\begin{center}\vspace{-2.5mm}
\includegraphics[width=8.6cm]{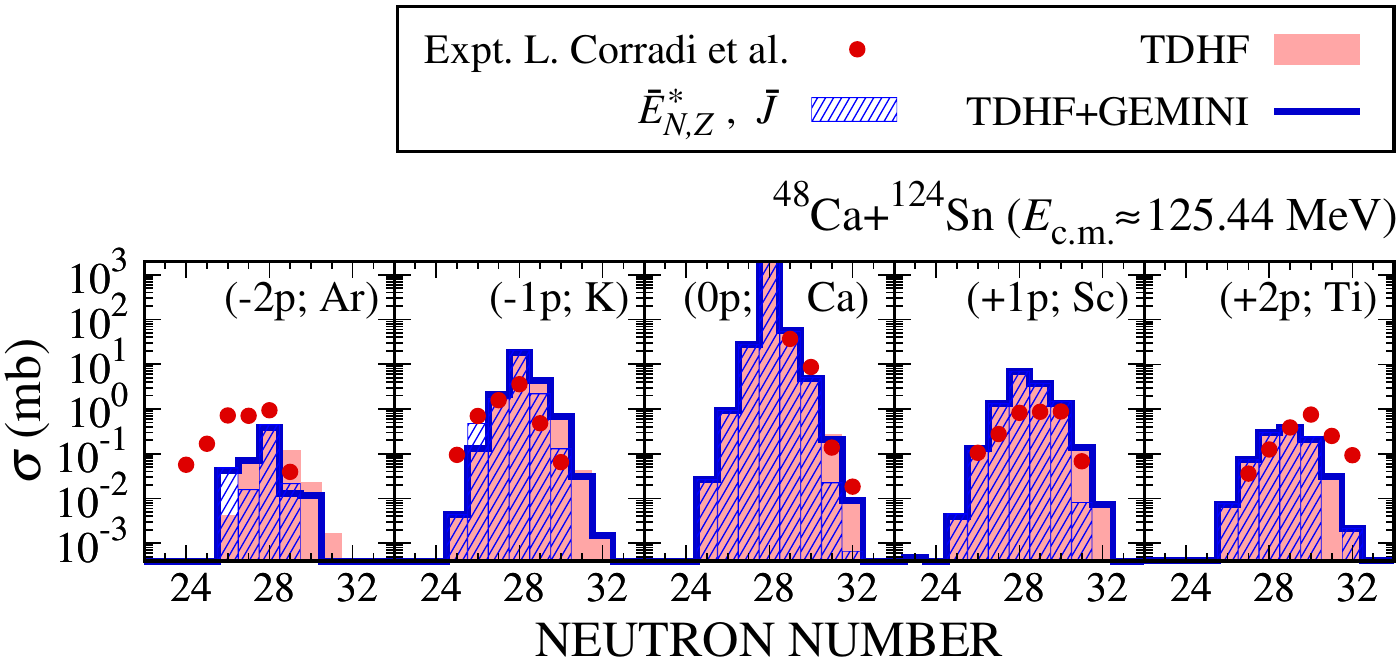}\vspace{-1mm}
\caption{
   Same as Fig.~\ref{FIG:NTCS_40Ca+124Sn_ave}, but for the $^{48}$Ca+$^{124}$Sn
   reaction at $E_{\rm c.m.}\simeq125.44$~MeV.
}
\label{FIG:NTCS_48Ca+124Sn_ave}
\end{center}
\end{figure}

\begin{figure}[t]
\begin{center}
\includegraphics[width=8.6cm]{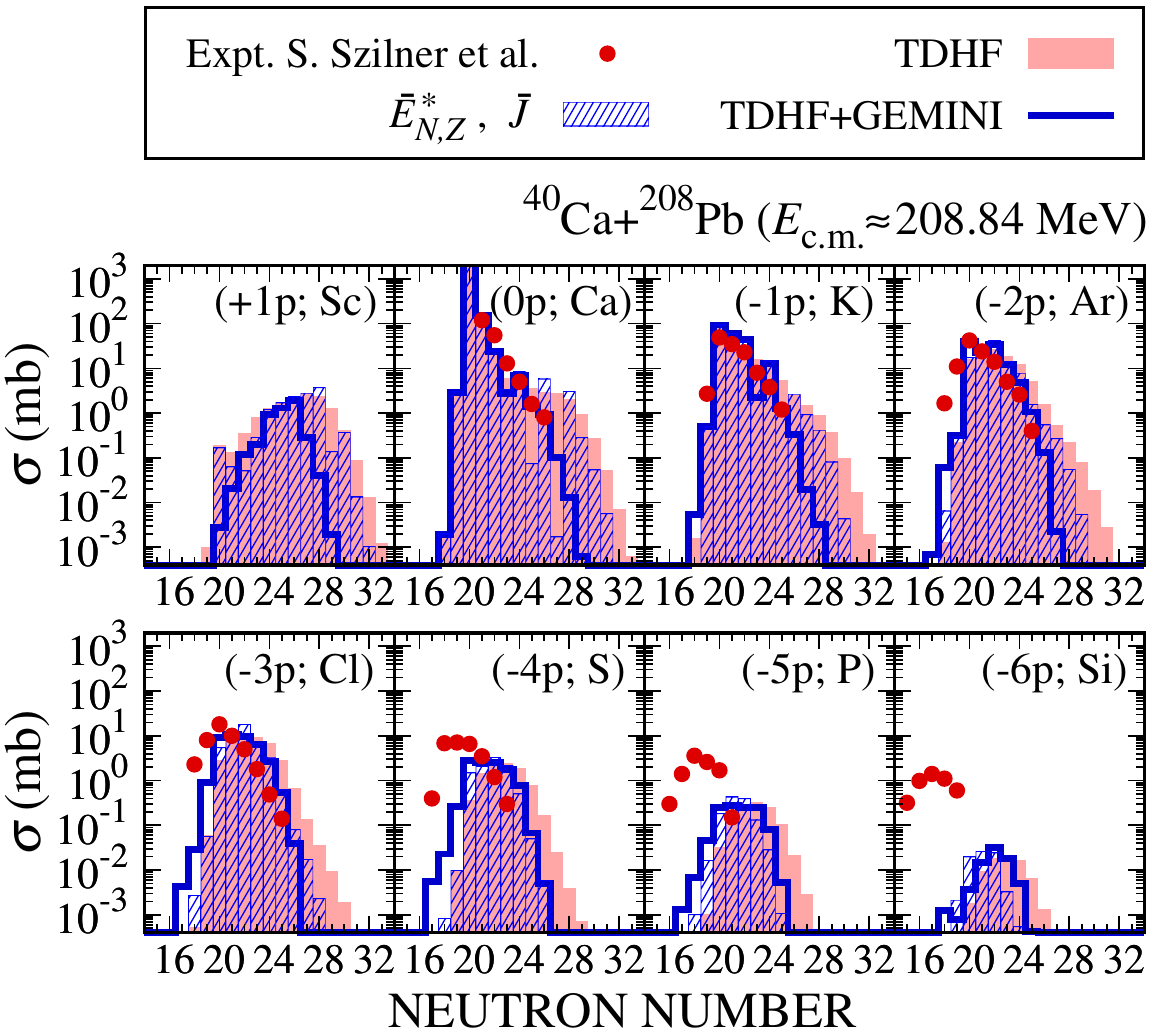}\vspace{-1mm}
\caption{
   Same as Figs.~\ref{FIG:NTCS_40Ca+124Sn_ave} and \ref{FIG:NTCS_48Ca+124Sn_ave},
   but for the $^{40}$Ca+$^{208}$Pb reaction at $E_{\rm c.m.}\simeq208.84$~MeV.
}\vspace{-7.5mm}
\label{FIG:NTCS_40Ca+208Pb_E249_ave}
\end{center}
\end{figure}

\begin{figure}[b]
\begin{center}\vspace{-2.5mm}
\includegraphics[width=8.6cm]{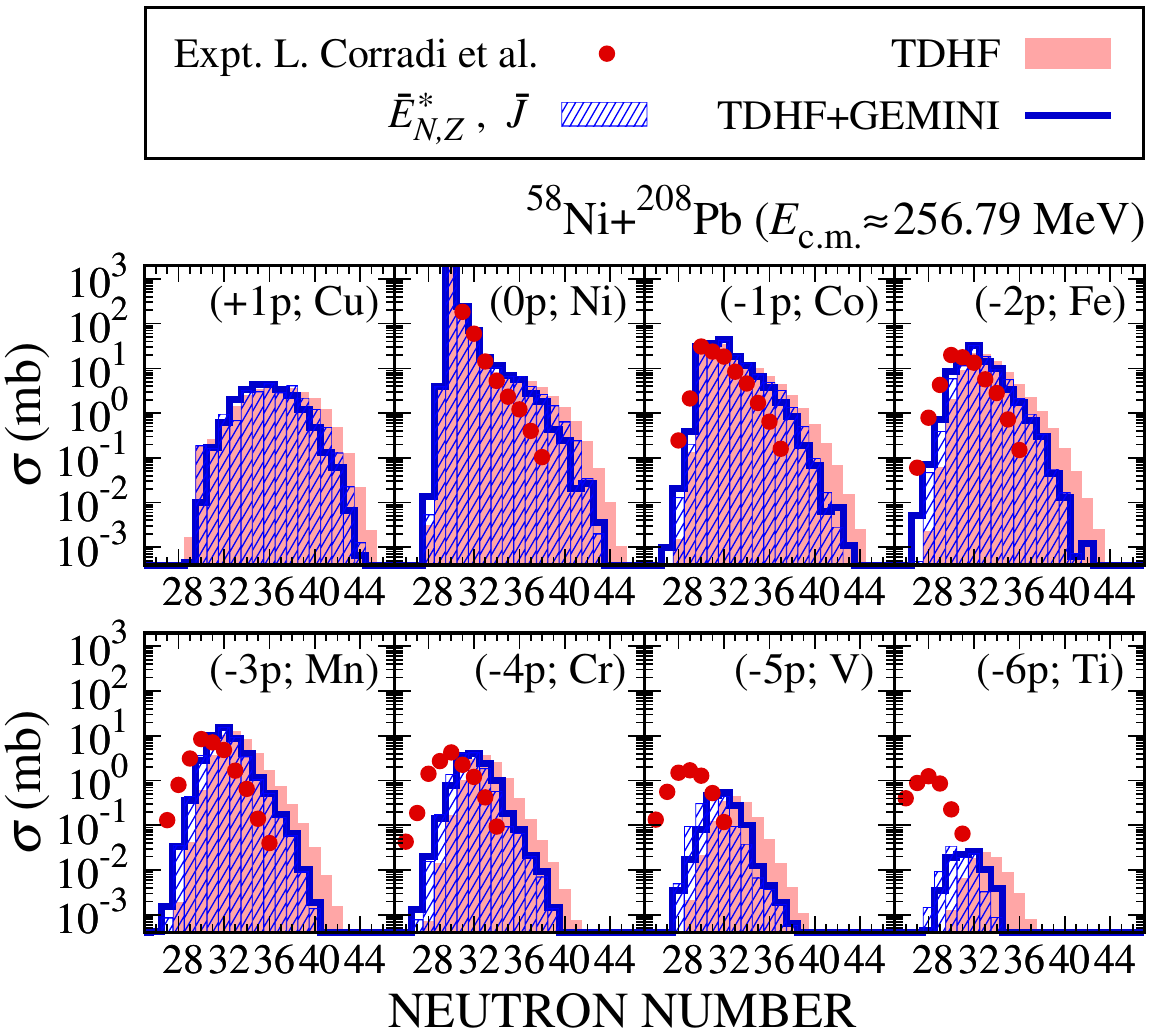}\vspace{-1mm}
\caption{
   Same as Figs.~\ref{FIG:NTCS_40Ca+124Sn_ave}--\ref{FIG:NTCS_40Ca+208Pb_E249_ave},
   but for the $^{58}$Ni+$^{208}$Pb reaction at $E_{\rm c.m.}\simeq256.79$~MeV.
}
\label{FIG:NTCS_58Ni+208Pb_ave}
\end{center}
\end{figure}

The results of TDHF+GEMINI calculations with the average quantities, $\bar{E}_{N,Z}^*$
and $\bar{J}$, are presented in Figs.~\ref{FIG:NTCS_40Ca+124Sn_ave}--\ref{FIG:NTCS_58Ni+208Pb_ave}
by blue shaded histograms. As seen in the figures, this treatment provides quantitatively similar results as those obtained
with the elaborated PNP calculations (shown in Figs.~\ref{FIG:NTCS_40Ca+124Sn}--\ref{FIG:NTCS_58Ni+208Pb}).
For some channels, evaporation effects are less pronounced (\textit{e.g.} for the $^{40}$Ca+$^{124}$Sn reaction
shown in Fig.~\ref{FIG:NTCS_40Ca+124Sn_ave}), which might be due to the assumption that the excitation energy
is shared as it is proportional to fragment masses. On the other hand, the angular momentum effects on neutron
evaporation processes turned out to be negligibly small in the reactions investigated. It may play a certain role
in fission processes of heavier fragments, and the validity of using $\bar{J}$ should be re-examined for such cases.
Nevertheless, the results shown in Figs.~\ref{FIG:NTCS_40Ca+124Sn}--\ref{FIG:NTCS_58Ni+208Pb_ave} may
support the usage of the average values for the evaluation of secondary deexcitation processes, significantly
reducing the computational cost. In the following, TDHF+GEMINI is applied to the $^{64}$Ni+$^{238}$U and
$^{136}$Xe+$^{198}$Pt reactions employing the simpler treatment with the average values, $\bar{E}_{N,Z}^*$
and $\bar{J}$, for statistical-model calculations (We will keep the usage of Eq.~(\ref{Eq:Eex_TKEL}) as a simple,
conservative choice).

\begin{figure*}[t]
   \begin{center}
   \includegraphics[width=16.5cm]{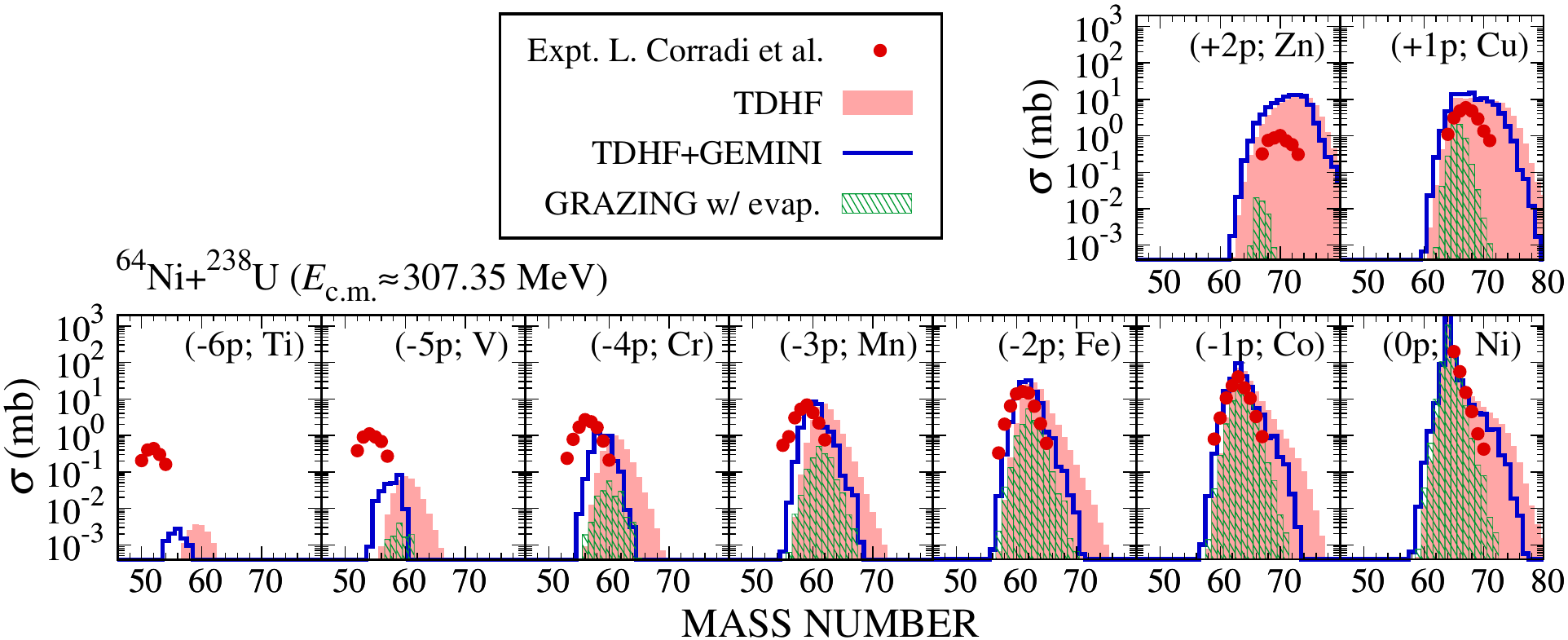}
   \end{center}\vspace{-2mm}
   \caption{
   Production cross sections for lighter fragments in the $^{64}$Ni+$^{238}$U reaction at
   $E_{\rm c.m.}\simeq307.35$~MeV. Each panel shows cross sections for different proton-transfer channel.
   The horizontal axis is the mass number of the fragments. Cross sections associated with side collisions are shown (see text).
   Cross sections for secondary products evaluated by TDHF+GEMINI (with $\bar{E}_{N,Z}^*$ and $\bar{J}$ as in
   Figs.~\ref{FIG:NTCS_40Ca+124Sn_ave}--\ref{FIG:NTCS_58Ni+208Pb_ave}) are shown by blue solid lines.
   The experimental data were reported in Ref.~\cite{Corradi(64Ni+238U)}. \fs{GRAZING} results \cite{GRAZING-online}
   are also shown by green shaded histograms, for comparison.
   }\vspace{-1mm}
   \label{FIG:NTCS_64Ni+238U}
\end{figure*}

Figure~\ref{FIG:NTCS_64Ni+238U} shows production cross sections for the $^{64}$Ni+$^{238}$U
reaction at $E_{\rm c.m.}\simeq307.35$~MeV. Although $^{238}$U is largely deformed in a prolate
shape, it has been shown that MNT processes in peripheral collisions do not depend much on the nuclear
orientations \cite{KS_KY_Ni-U}. Here, the results of side collisions ($z$-direction case in Ref.~\cite{KS_KY_Ni-U}),
where the symmetry axis of $^{238}$U is always set perpendicular to the reaction plane, are shown for
better visibility. (Other cases resulted in quantitatively similar cross sections.) Since the $^{64}$Ni+$^{238}$U
reaction has a relatively large $N/Z$ asymmetry, neutron-pickup and proton-stripping processes dominate
the nucleon transfer. The agreement and disagreement are indeed similar to the $^{40}$Ca+$^{124}$Sn,
$^{40}$Ca+$^{208}$Pb, and $^{58}$Ni+$^{208}$Pb cases shown in
Figs.~\ref{FIG:NTCS_40Ca+124Sn},\,\ref{FIG:NTCS_40Ca+208Pb_E249},\,\ref{FIG:NTCS_58Ni+208Pb} (and
\ref{FIG:NTCS_40Ca+124Sn_ave},\,\ref{FIG:NTCS_40Ca+208Pb_E249_ave},\,\ref{FIG:NTCS_58Ni+208Pb_ave}).
A characteristic feature specific to the $^{64}$Ni+$^{238}$U reaction is that cross sections were measured
also for proton-pickup processes, ($+1p$) and ($+2p$). While \fs{GRAZING} substantially underestimates those
cross sections, especially for the ($+2p$) channel, TDHF+GEMINI provides significant cross sections, even greater
than the experimental data. In TDHF, the latter originated from a transitional regime from quasielastic to more
complicated reactions, like deep-inelastic and QF processes. Indeed, if we exclude contributions from small
impact parameters ($b\lesssim4$~fm), where energy loss is already saturated \cite{KS_KY_Ni-U}, the
overestimation of cross sections for proton-pickup processes can be removed, leaving cross sections for
proton-stripping processes almost unaffected. Similarly, the ``shoulders" due to the overestimation of
neutron-pickup processes seen in Figs.~\ref{FIG:NTCS_40Ca+208Pb_E249}~and~\ref{FIG:NTCS_58Ni+208Pb}
for the $^{40}$Ca+$^{208}$Pb and $^{58}$Ni+$^{208}$Pb reactions can be removed if one excludes
contributions from small impact parameters, where the onset of mass equilibration process through a thick-neck
formation and its breaking is observed \cite{KS_KY_MNT}. However, all figures in the present paper show
cross sections obtained by integration over the full impact-parameter range, avoiding any \textit{ad hoc}
manipulation of the obtained results.

The primary goal of this work is to predict optimal conditions to produce new neutron-rich
unstable isotopes. Aiming at production of neutron-rich nuclei around the neutron magic
number $N=126$, whose properties are crucial to understand the detailed scenario
of the $r$-process nucleosynthesis, an experiment was recently carried out for the
$^{136}$Xe+$^{198}$Pt reaction \cite{Watanabe(136Xe+198Pt)}. In Ref.~\cite{Watanabe(136Xe+198Pt)},
production cross sections for heavier (target-like) fragments were deduced from detected
outcomes with respect to lighter (projectile-like) fragments, obtaining promising results.
Thus, the comparison with the experimental data for this system and TDHF+GEMINI is
expected to be a benchmark to check the accuracy and usefulness of the proposed method.

\begin{figure*}[t]
\begin{center}
\includegraphics[width=11.8cm]{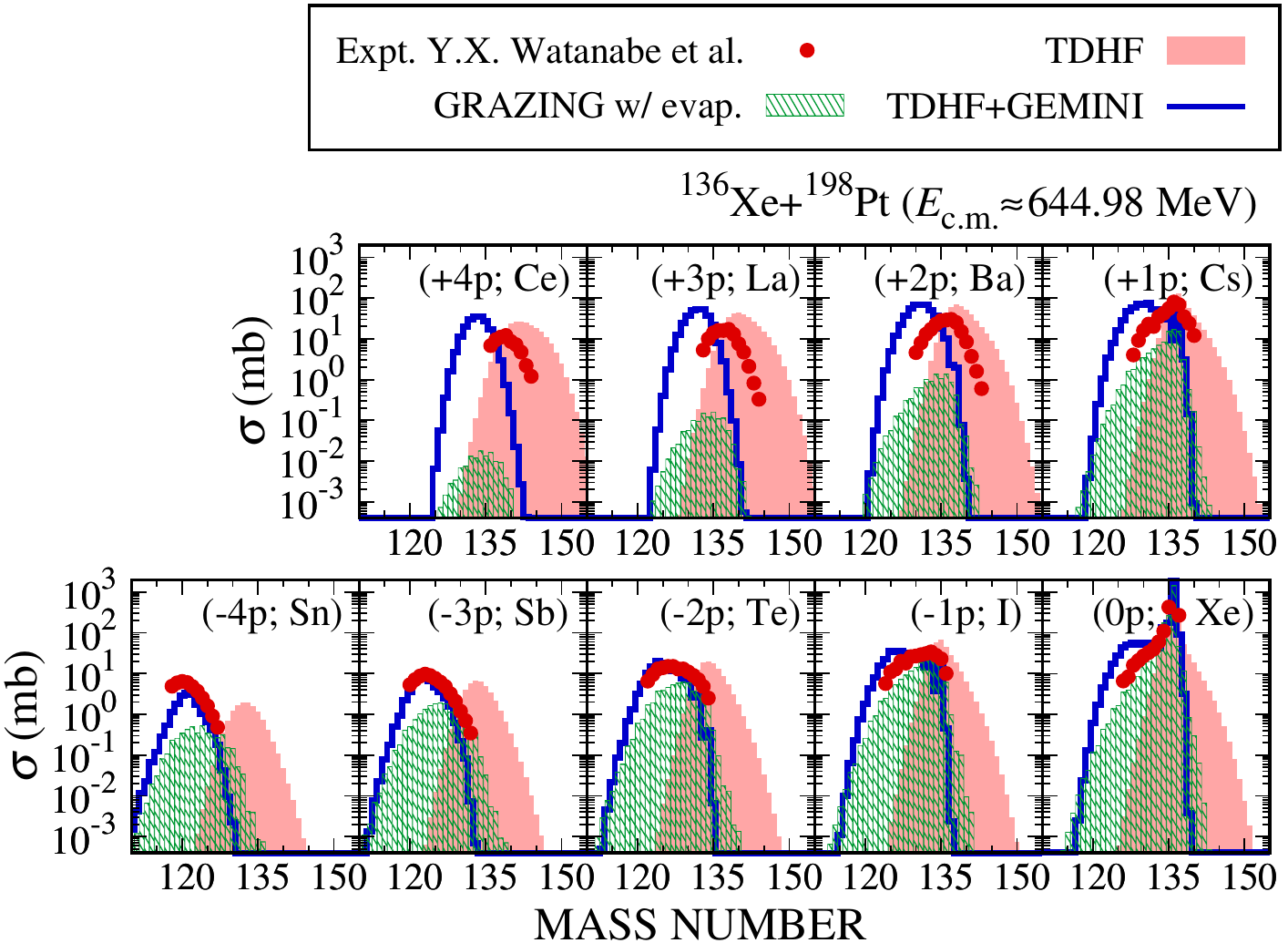}
\end{center}\vspace{-2mm}
\caption{
   Same as Fig.~\ref{FIG:NTCS_64Ni+238U}, but for the $^{136}$Xe+$^{198}$Pt reaction at
   $E_{\rm c.m.}\simeq644.98$~MeV. The experimental data and the \fs{GRAZING} results
   were taken from Ref.~\cite{Watanabe(136Xe+198Pt)} (see \cite{Watanabe} for a comment
   on this point).
   }\vspace{-1mm}
\label{FIG:NTCS_136Xe+198Pt}
\end{figure*}

In Fig.~\ref{FIG:NTCS_136Xe+198Pt}, production cross sections for the $^{136}$Xe+$^{198}$Pt
reaction at $E_{\rm c.m.}\simeq644.98$~MeV are shown. It should be noted here that the projectile
and the target have very similar $N/Z$ ratios (\textit{cf.}~Table~\ref{table}), and one expects transfer
of neutrons and protons toward both directions (similar to the $^{48}$Ca+$^{124}$Sn reaction shown
in Figs.~\ref{FIG:NTCS_48Ca+124Sn} and \ref{FIG:NTCS_48Ca+124Sn_ave}). By comparing the cross
sections for primary (red filled areas) and secondary (blue solid lines) reaction products, one can see significant
effects of deexcitation processes. For proton-stripping channels ($-xp$), TDHF+GEMINI reproduces the
measurements surprisingly well, both the magnitude and the centroid of the cross sections for the secondary
products. On the other hand, too large deexcitation effects are observed for proton-pickup channels ($+xp$).
This type of disagreement is peculiar to the $^{136}$Xe+$^{198}$Pt reaction. The experimental data
indicate that reaction products in proton-pickup channels might be less excited compared to those in
proton-stripping channels. Note that \fs{GRAZING} also provides similar magnitude of evaporation effects,
although the absolute value of the cross sections are substantially underestimated.

Lastly, it is to be reminded that TDHF+GEMINI also allows to evaluate production
cross sections for heavier fragments, where transfer-induced fission is expected
to play an important role. Detailed investigation of production mechanisms of heavy
neutron-rich nuclei which are survived against transfer-induced fission as well as
particle evaporation is the next step of this work.

\section{SUMMARY}{\label{Sec:summary}}

In this paper, a method, called TDHF+GEMINI, has been proposed, which
enables us to evaluate production cross sections for \textit{secondary} products
in low-energy heavy ion reactions. In the method, the reaction dynamics, on the
timescale of $10^{-21}$--$10^{-20}$~sec, is described microscopically based
on the time-dependent Hartree-Fock (TDHF) theory. Production probabilities,
total angular momenta, and excitation energies of primary reaction products are
extracted from the TDHF wavefunction after collision, using the particle-number
projection method. Based on those quantities derived from TDHF, secondary
deexcitation processes of primary reaction products, including both particle
evaporation and fission, are described employing the \fs{GEMINI}++
compound-nucleus deexcitation model.

The method was applied to $^{40,48}$Ca+$^{124}$Sn, $^{40}$Ca+$^{208}$Pb,
$^{58}$Ni+$^{208}$Pb, $^{64}$Ni+$^{238}$U, and $^{136}$Xe+$^{198}$Pt
reactions for which precisely-measured experimental cross sections are available. The
inclusion of deexcitation effects, which are dominated by neutron evaporation in the
present cases, changes the cross sections toward the direction consistent with the
experimental data. However, there still remain discrepancies between the measured
cross sections and the TDHF+GEMINI results, especially for multi-proton transfer processes.
It may indicate the importance of description going beyond the standard self-consistent
mean-field theory to correctly describe multinucleon transfer processes in low-energy
heavy ion reactions.

In conclusion, it has been demonstrated that, even though some discrepancy still remains,
the combination of TDHF and a statistical model offers an excellent starting point toward a complete
modeling of low-energy heavy ion reactions. It is important to stress that, in the proposed method,
there is no room to adjust the model parameters specific to each reaction: energy density functional
is determined so as to reproduce known properties of finite nuclei and nuclear matter \cite{Chabanat};
\fs{GEMINI}++ \cite{GEMINI++} and its ongoing developments \cite{Charity2010,Mancusi2010}
allow a systematic reproduction of a large body of data for the entire mass region. Therefore, it will
be a promising tool that can predict, in a non-empirical way, optimal reaction mechanisms to produce
new neutron-rich isotopes that have not yet been produced to date.

\begin{acknowledgments}
The author wish to thank Kazuhiro Yabana (University of Tsukuba) for useful comments on this article.
The author is grateful to Lorenzo Corradi (INFN-LNL) and Yutaka Watanabe (KEK-Japan)
for providing the experimental data. Lu Guo (University of Chinese Academy of Sciences)
is also thanked for useful information on \fs{GEMINI}++. The author acknowledges support of
Polish National Science Centre (NCN) Grant, decision No.~DEC-2013/08/A/ST3/00708. This work is
indebted for continuous usage of computational resources of the HPCI system (HITACHI SR16000/M1)
provided by Information Initiative Center (IIC), Hokkaido University, through the HPCI System Research
Projects (Project IDs: hp120204, hp140010, 150081, hp160062, and hp170007).
\end{acknowledgments}

\end{document}